\documentclass[prl,twocolumn]{revtex4}
\usepackage{graphicx}

\begin{document}

\title{Real-time detection of single electron tunneling using a quantum point contact}

\author{L.M.K. Vandersypen, J.M. Elzerman, R.N. Schouten, L.H. Willems van Beveren, R. Hanson and L.P. Kouwenhoven}
\affiliation{Kavli Institute of NanoScience and ERATO Mesoscopic
Correlation Project, Delft University of Technology, Lorentzweg 1,
2628 CJ Delft, The Netherlands}

\date{\today}

\def\be{\begin{equation}}
\def\ee{\end{equation}}
\newcommand{\ket}[1]{\mbox{$|#1\rangle$}}

\def\rHz{$\sqrt{\mbox{Hz}}$}

\begin{abstract}
We observe individual tunnel events of a single electron between a quantum dot and a reservoir, using a nearby quantum point contact (QPC) as a charge meter. The QPC is capacitively coupled to the dot, and the QPC conductance changes by about 1$\%$ if the number of electrons on the dot changes by one. The QPC is voltage biased and the current is monitored with an IV-convertor at room temperature. We can resolve tunnel events separated by only 8 $\mu$s, limited by noise from the IV-convertor. Shot noise in the QPC sets a 25 ns lower bound on the accessible timescales.
\end{abstract}

\maketitle

Fast and sensitive detection of charge has greatly propelled the study of single-electron phenomena. The most sensitive electrometer known today is the single-electron transistor (SET)~\cite{Fulton87a}, incorporated into a radio-frequency resonant circuit~\cite{Schoelkopf98a}. Such RF-SETs can be used for instance to detect charge fluctuations on a quantum dot, capacitively coupled to the SET island~\cite{Lu03a,Fujisawa04a}. Already, real-time electron tunneling between a dot and a reservoir has been observed on a sub-$\mu$s timescale~\cite{Lu03a}.

A much simpler electrometer is the quantum point contact (QPC). The conductance, $G_{Q}$, through the QPC channel is quantized, and at the transitions between quantized conductance plateaus, $G_{Q}$ is very sensitive to the electrostatic environment, including the number of electrons, $N$, on a dot in the vicinity~\cite{Field93a}. This property has been exploited to measure  fluctuations in $N$ in real-time, on a timescale from seconds~\cite{Cooper00a} down to about 10 ms~\cite{Schleser04a}.

Here we demonstrate that a QPC can be used to detect single-electron charge fluctuations in a quantum dot in less than $10\, \mu$s, and analyze the fundamental and practical limitations on sensitivity and bandwidth. 

\begin{figure}
\begin{center}
\includegraphics[width=8.5cm]{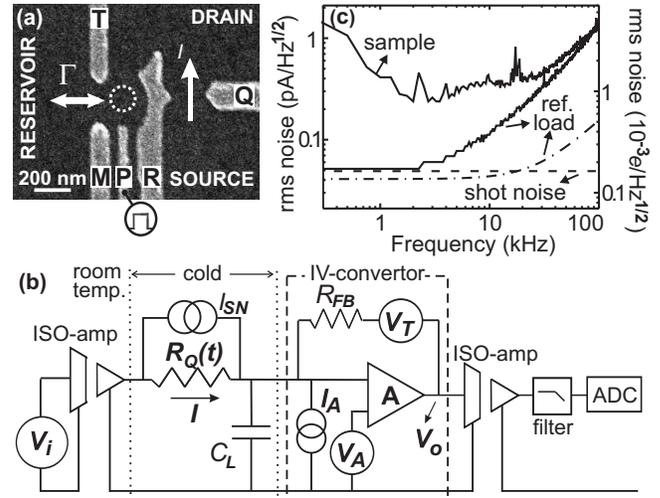}
\end{center}
\vspace*{-.5cm}
\caption{(a) Scanning electron micrograph of a device as used in the experiment (gates which are grounded are hidden). Gates $T, M$ and $R$ define the quantum dot (dotted circle), and gates $R$ and $Q$ form the QPC. Gate $P$ is connected to a pulse source via a coaxial cable. See~\cite{Elzerman03a} for a more detailed description. 
(b) Schematic of the experimental set-up, including the most relevant noise sources. The QPC is represented by a resistor, $R_Q$. (c) Noise spectra measured when the IV-convertor is connected to the sample (top solid trace), and, for reference, to an open-ended 1 m twisted pair of wires (lower solid trace). The latter represents a 300 pF load, if we include the 200 pF measured amplifier input capacitance. The diagram also shows the calculated noise level for the 300 pF reference load, neglecting $I_A$ (dotted-dashed), and the shot noise limit (dashed). The left and right axes express the noise in terms of current through the QPC and electron charge on the dot respectively.}
\vspace*{-.6cm}
\label{fig:device}
\end{figure}

The quantum dot and QPC are defined in the two-dimensional electron gas (2DEG) formed at a GaAs/Al$_{0.27}$Ga$_{0.73}$As interface 90 nm below the surface, by applying negative voltages to metal surface gates (Fig.~\ref{fig:device}a). The device is attached to the mixing chamber of a dilution refrigerator with a base temperature of 20 mK, and the electron temperature is $\sim$ 300 mK in this measurement. The dot is set near the $N=0$ to $N=1$ transition, with the gate voltages tuned such that the dot is isolated from the QPC drain, and has a small tunnel rate, $\Gamma$, to the reservoir. Furthermore, the QPC conductance is set at $G_Q = 1/R_Q \approx (30 \,$k$\Omega)^{-1}$, roughly halfway the transition between $G_Q=2e^2/h$ and $G_Q=0$, where it is most sensitive to the electrostatic environment~\cite{spinsplit}.

A schematic of the electrical circuit is shown in Fig.~\ref{fig:device}b. The QPC source and drain are connected to room temperature electronics by signal wires, which run through Cu-powder filters at the mixing chamber to block high frequency noise ($> 100$ MHz) coming from room temperature. Each signal wire is twisted with a ground wire from room temperature to the mixing chamber. A voltage, $V_i$, is applied to the source via a home-built opto-coupled isolation stage. The current through the QPC, $I$, is measured via an IV-convertor connected to the drain, and an opto-coupled isolation amplifier, both home-built as well. The IV-convertor is based on a dual low-noise JFET (Interfet 3602). Finally, the signal is AC-coupled to an 8th-order elliptic low-pass filter (SRS650), and the current fluctuations, $\Delta I$, are digitized at $2.2 \times 10^6$ 14-bit samples per second (ADwin Gold).

The measurement bandwidth is limited by the low-pass filter formed by the capacitance of the line and Cu-powder filters, $C_L \approx 1.5$ nF, and the input impedance of the IV-convertor, $R_i = R_{FB} / A$. Thermal noise considerations (below) impose $R_{FB}=10\,$M$\Omega$. We choose the amplifier gain $A = 10000$, such that $1 / (2\pi R_i C_L) \approx 100$ kHz~\cite{bandwidth}. However, we shall see that the true limitation to measurement speed is not the bandwidth but the signal-to-noise ratio as a function of frequency.

The measured signal corresponding to a single electron charge leaving the dot amounts to $\Delta I \approx 0.3$ nA with the QPC biased at $V_i = 1$ mV, a $1\%$ change in the overall current $I$ ($I \approx 30$ nA, consistent with the series resistance of $R_Q$, $R_i=1$ k$\Omega$ and the resistance of the Ohmic contacts of about $2$ k$\Omega$). Naturally, the signal strength is proportional to $V_i$, but we found that for $V_i \ge 1$ mV, the dot occupation was affected, possibly due to heating. We therefore proceed with the analysis using $I = 30$ nA and $\Delta I = 0.3$ nA.

The most relevant noise sources~\cite{Horowitz89a} are indicated in the schematic of Fig.~\ref{fig:device}b. In Table~\ref{tab:noise}, we give an expression and value for each noise contribution in terms of rms current at the IV-convertor input, so it can be compared directly to the signal, $\Delta I$. We also give the corresponding value for the rms charge noise on the quantum dot. Shot noise, $I_{SN}$, is intrinsic to the QPC and therefore unavoidable. Both $I_{SN}$ and $\Delta I$ are zero at QPC transmission $T=0$ or $T=1$, and maximal at $T=1/2$; here we use $T \le 1/2$. The effect of thermal noise, $V_T$, can be kept small compared to other noise sources by choosing $R_{FB}$ sufficiently large; here $R_{FB} = 10$ M$\Omega$. The JFET input voltage noise is measured to be $V_A = 0.8$ nV/\rHz. As a result of $V_A$, a noise current flows from the IV-convertor input leg to ground, through the QPC in parallel with the line capacitance. Due to the capacitance, $C_L$, the rms noise current resulting from $V_A$ increases with frequency; it equals $\Delta I$ at 120 kHz. There is no specification available for the JFET input current noise, $I_A$, but for comparable JFETs, $I_A$ is a few fA/\rHz$\,$ at 1 kHz.

\begin{table}
\begin{tabular}{c|c|c|c}
noise          & \multicolumn{2}{c|}{rms noise current}     & rms charge noise\\ 
source         & expression                            & A/\rHz    & $e$/\rHz \\ \hline
$I_{SN}$       & $\sqrt{T(1-T) 2eI}$                   &$49\times10^{-15}$&$1.6\times10^{-4}$\\
$V_T$          & $\sqrt{4 k_B T / R_{FB}}$             &$41\times10^{-15}$&$1.4\times10^{-4}$\\
$V_A$          & $V_A \frac{1 + j2\pi f R_Q C_L}{R_Q}$ & \\
$V_A$, low $f$ & $V_A/R_{FB}$                          &$32\times10^{-15}$&$1.1\times10^{-4}$\\
$V_A$, high $f$& $V_A 2\pi f C_L$                  &$7.5\times10^{-18}f$&$2.5 \times 10^{-8} f $\\
$I_A$          & $I_A$                                 &  -      & -
\end{tabular}
\caption{Contributions to the noise current at the IV-convertor input. By dividing the noise current by 300 pA (the signal corresponding to one electron charge leaving the dot), we obtain the rms charge noise on the dot.}
\vspace*{-.3cm}
\label{tab:noise}
\end{table}

We summarize the expected noise spectrum in Fig.~\ref{fig:device}c, and compare this with the measured noise spectrum in the same figure. For a capacitive reference load $C_L=300$ pF, the noise level measured below a few kHz is 52 fA/\rHz, close to the noise current due to $V_T$, as expected; at high frequencies, the measured noise level is significantly higher than would be caused by $V_A$ in combination with the 300 pF load, so it appears that $I_A$ rapidly increases with frequency. With the sample connected, we observe substantial $1/f^2$ noise ($1/f$ in the noise amplitude), presumably from spurious charge fluctuations near the QPC, as well as interference at various frequencies. Near 100 kHz, the spectrum starts to roll off because of the 100 kHz low-pass filter formed by $C_L = 1.5$ nF and $R_i = 1 $ k$\Omega$ (for the reference load, $C_L$ is only $300$ pF so the filter cut-off is at 500 kHz). 

From the data, we find that the measured charge noise integrated from DC is about three times smaller than $e$ at 40 kHz. We set the cut-off frequency of the external low-pass filter at 40 kHz, so we should see clear steps in time traces of the QPC current, corresponding to single electrons tunneling on or off the dot.

\begin{figure}
\begin{center}
\includegraphics[width=8.5cm]{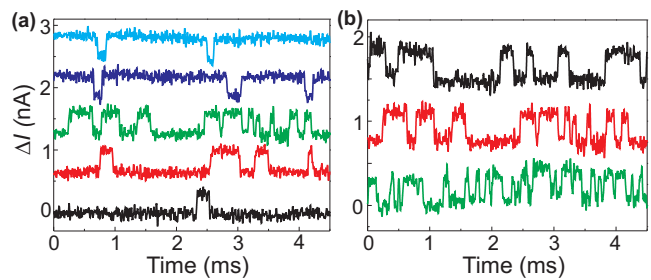}
\end{center}
\vspace*{-.7cm}
\caption{Measured changes in the QPC current, $\Delta I$, with the electrochemical potential in the dot and in the reservoir nearly equal. $\Delta I$ is ``high'' and ``low'' for $0$ and $1$ electrons on the dot respectively ($V_i = 1$ mV; the steps in $\Delta I$ are $\approx 0.3$ nA). Traces are offset for clarity. (a) The dot potential is lowered from top to bottom. (b) The tunnel barrier is lowered from top to bottom.}
\vspace*{-.5cm}
\label{fig:rts}
\end{figure}

We test this experimentally, in the regime where the electrochemical potential in the dot is nearly lined up with the electrochemical potential in the reservoir. The electron can then {\em spontaneously} tunnel back and forth between the dot and the reservoir, and the QPC current should exhibit a random telegraph signal (RTS). This is indeed what we observe experimentally (Fig.~\ref{fig:rts}). In order to ascertain that the RTS really originates from electron tunnel events between the dot and the reservoir, we verify that (1) the dot potential relative to the Fermi level determines the fraction of the time an electron resides in the dot (Fig.~\ref{fig:rts}a) and (2) the dot-reservoir tunnel barrier sets the RTS frequency (Fig.~\ref{fig:rts}b). The rms baseline noise is $~\sim 0.05$ nA and the shortest steps that clearly reach above the noise level are about $8\, \mu$s long. This is consistent with the 40 kHz filter frequency, which permits a rise time of $8\, \mu$s.

\begin{figure}
\begin{center}
\includegraphics[width=8.5cm]{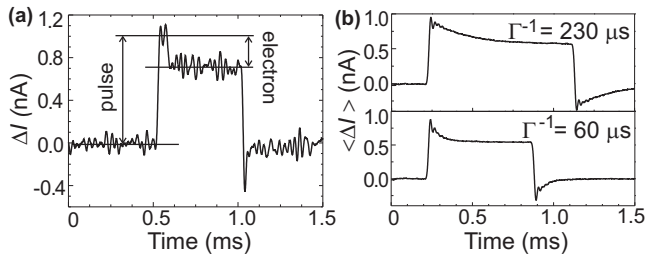}
\end{center}
\vspace*{-.7cm}
\caption{(a) Measured changes in the QPC current, $\Delta I$, when a pulse is applied to gate $P$, near the degeneracy point between $0$ and $1$ electrons on the dot ($V_i = 1$ mV). (b) Average of 286 traces as in (a). The top and bottom panel are taken with a different setting of gate $M$. The damped oscillation following the pulse edges is due to the 8th-order 40 kHz filter.}
\vspace*{-.4cm}
\label{fig:pulse}
\end{figure}

Next, we {\em induce} tunnel events by pulsing the dot potential, so $N$ predictably changes from $0$ to $1$ and back to $0$. The response of the QPC current to such a pulse contains two contributions (Fig.~\ref{fig:pulse}a). First, the shape of the pulse is reflected in $\Delta I$, as the pulse gate couples capacitively to the QPC. Second, some time after the pulse is started, an electron tunnels into the dot and $\Delta I$ goes down by about 0.3 nA. Similarly, $\Delta I$ goes up by 0.3 nA when an electron leaves the dot, some time after the pulse ends. We observe that the time before tunneling takes place is randomly distributed, and obtain a histogram of this time simply by averaging over many single-shot traces (Fig.~\ref{fig:pulse}b). The measured distribution decays exponentially with the tunnel time, characteristic of a Poisson process. The average time before tunneling corresponds to $\Gamma^{-1}$, and can be tuned by adjusting the tunnel barrier.

Our measurements clearly demonstrate that a QPC can serve as a fast and sensitive charge detector. Compared to an SET, a QPC offers several practical advantages. First, a QPC requires fabrication and tuning of just a single additional gate when integrated with a quantum dot defined by metal gates, whereas an SET requires two tunnel barriers, and a gate to set the island potential. Second, QPCs are more robust and easy to use in the sense that spurious, low-frequency fluctuations of the electrostatic potential hardly change the QPC sensitivity to charges on the dot (the transition between quantized conductance plateaus has an almost constant slope over a wide range of electrostatic potential), but can easily spoil the SET sensitivity.

With an RF-SET, a sensitivity to charges on a quantum dot of $\sim 2 \times 10^{-4} e/$\rHz$\,$ has been reached over a 1 MHz bandwidth~\cite{Lu03a}. Theoretically even a ten times better sensitivity is possible~\cite{Schoelkopf98a}. Could a QPC perform equally well?

The noise level in the present measurement could be reduced by a factor of two to three using a JFET input-stage which better balances input voltage noise and input current noise. Further improvements can be obtained by lowering $C_L$, either by reducing the filter capacitance, or by placing the IV-convertor closer to the sample, inside the refrigerator. The bandwidth would also increase as it is inversely proportional to $C_L$.

Much more significant reductions in the instrumentation noise could be realized by embedding the QPC in a resonant electrical circuit and measuring the damping of the resonator, analogous to the operation of an RF-SET. We estimate that with an ``RF-QPC'' and a low-temperature HEMT amplifier, the sensitivity could be $2 \times 10^{-4}e/$\rHz$\,$. At this point, the noise current from the amplifier circuitry is comparable to the QPC shot noise. Furthermore, the bandwidth does not depend on $C_L$ in reflection measurements, and can easily be 1 MHz.

To what extent the signal can be increased is unclear, as we do not yet understand the mechanism through which the dot occupancy is disturbed for $V_i>1$ mV~\cite{backaction}. Certainly, the capacitive coupling of the dot to the QPC channel can easily be made five times larger than it is now by optimizing the gate design~\cite{Cooper00a}. Keeping $V_i=1$ mV, the sensitivity would then be $4 \times 10^{-5}e/$\rHz$\,$.

Finally, we point out that, unlike a SET, a QPC can reach the quantum limit of detection~\cite{qlimit}, where the measurement induced decoherence takes the minimum value permitted by quantum mechanics. Qualitatively, this is because (1) information on the charge state of the dot is transferred only to the QPC current and not to degrees of freedom which are not observed, and (2) an external perturbation in the QPC current does not couple back to the charge state of the dot.

We thank R. Schoelkopf, K. Schwab, K. Harmans and L. Saminadayar for useful discussions, T. Fujisawa T. Hayashi, T. Saku and Y. Hirayama for help with device fabrication, and the DARPA-QUIST program, the ONR, the EU-RTN network on spintronics, and the Dutch Organisation for Fundamental Research on Matter (FOM) for financial support.


\begin{thebibliography}{30}
\bibitem{spinsplit} Despite a $B=10$ T field in the plane of the 2DEG, no spin-split plateau in the QPC conductance is visible in this sample.
\bibitem{bandwidth} The bandwidth of the amplifier inside the IV-convertor is 500 kHz, and the output ISO-amp bandwidth is 300 kHz.
\bibitem{backaction} The statistics of the RTS was altered for $V_i>1$ mV, irrespective of (1) whether $V_i$ was applied to the QPC source or drain, (2) the potential difference between the reservoir and the QPC source/drain, and (3) the QPC transmission $T$.
\end{thebibliography}
\end{document}